# Te Doped Indium (II) Selenide Photocatalyst for Water Splitting: A First Principles Study

M.R. Ashwin Kishore[1] and P. Ravindran[1,2,*]

[1]Department of Physics, Central University of Tamil Nadu, Thiruvarur, Tamil Nadu, 610101, India
[2]Center of Material Science and Nanotechnology and Department of Chemistry, University of Oslo, Box 1033 Blindern, N-0315 Oslo, Norway    email: raviphy@cutn.ac.in

**Abstract.** Graphene like 2D materials have attracted tremendous attention in the field of photocatalytic water splitting. Indium Selenide (InSe) is one such potential material. Here, we report the effect of Tellurium on InSe monolayer for the photocatalytic water splitting by means of density functional calculations. The calculated bandgaps fall within the visible region of the solar spectrum indicating these materials could absorb a significant amount of solar light. Density of states calculations show that the covalent character is present between the atoms which is typical for the layered system. The band alignment with respect to redox potentials show that Te doped InSe is more favorable for the hydrogen reduction reaction than the pristine InSe monolayer. Our overall results show that Te substitution improves the photocatalytic water splitting ability in InSe monolayer.

## INTRODUCTION

A range of two-dimensional (2D) semiconducting materials have been studied extensively after the experimental realization of graphene[1] since its zero bandgap limits its potential use in optoelectronics, energy conversion applications, etc. Photocatalytic water splitting to produce hydrogen with help of semiconductor and solar energy has drawn considerable attention for the sustainable energy production. 2D materials show great potential in photocatalytic water splitting due to high surface area available for photocatalytic reactions to occur. Also, the 2D nature reduces the migration path for the photogenerated charge carriers to reach the reactive sites. Hence, the possibility of recombination of these photogenerated carriers is low. [2] Group III-VI monochalcogenide (MX, M = Ga and In, X = S, Se, and Te) is new to 2D family. Recently, monolayer and few-layer InSe [3] with an atomically flat surface have been successfully fabricated in experiment. These Group-III monochalcogenides have drawn considerable attention in the field of solar energy conversion and optoelectronics. [4] H. L. Zhuang *et al.* [5] reported the Group-III monochalcogenide photocatalysts for water splitting using density functional calculations. In this study, we report the effect of Te doping on the photocatalytic mechanism of InSe monolayer using accurate density functional theory calculations.

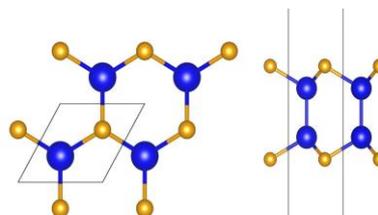

**FIGURE 1.** Top and side views of optimized InSe monolayer. The atoms are denoted as In (Blue) and Se (Gold) in this picture.

## COMPUTATIONAL DETAILS

All calculations were performed using projector augmented wave method implemented in VASP package. [6] The generalized gradient approximation proposed by Perdew, Burke and Ernzrhof (GGA-PBE) with van der Waals (vdW) correction proposed by Grimme was used for geometry optimization. For accurate band structure calculations, we employed Heyd-Scuseria-Ernzerhof (HSE06) hybrid functional with 25% Hartree-Fock exchange energy. Brillouin zone was sampled with a 9×9×1 Monkhorst-Pack **k**- point mesh for PBE functional and 3x3x1 **k**-mesh for more expensive HSE06 calculations. A cutoff energy of 400 eV for the plane wave basis set is used throughout all the calculations. A large vacuum of ~18 Å was used to avoid interaction between two adjacent periodic images. The force minimization steps were continued until the maximum Hellmann-Feynman forces acting on each atom is less than 0.01 eV/Å.

## RESULTS AND DISCUSSION

The crystal structure of InSe monolayer which exhibits hexagonal structure with $D_{3h}$ symmetry is shown in Fig.1. Our calculated lattice constant from GGA-PBE method ($a$ =4.08 Å) is consistent with previous theoretical [5] and experimental reports. [3] Owing to bandgap underestimation of PBE functional, we have calculated all the electronic properties using HSE06 functional. Results obtained from HSE06 functional show that the pristine and Te doped InSe monolayer are indirect bandgap semiconductors with valence band maxima (VBM) located between the Γ and M points and conduction band minima (CBM) located at the Γ point. (See Fig.2) The calculated bandgap of pristine and Te doped InSe is 2.14 and 2.24 eV, respectively. This increase in bandgap is due to the strain effect induced by Te doping. However, the calculated bandgaps fall within the visible region of the solar spectrum. Hence, these materials could harvest a significant amount of solar light which is crucial for photocatalytic water splitting. It may be noted that the direct bandgaps are close in size and position to indirect bandgaps with a maximum energy difference of 0.07 eV.

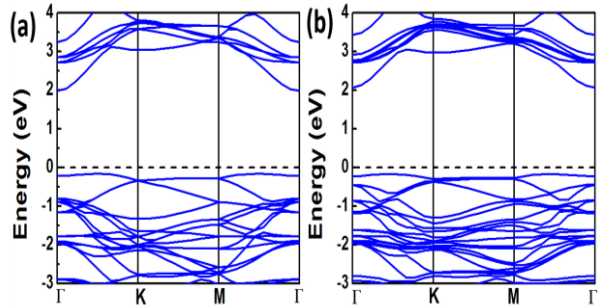

**FIGURE 2.** The calculated band structure of pristine InSe (a) and Te doped InSe (b) monolayer at the level of HSE06. The Fermi level is set to zero.

The calculated total and projected density of states for Te doped InSe monolayer is presented in Fig.3. It is found that the states at VBM is originating from the hybridization of Se-$p$, Te-$p$, and In-$p$ states while the CBM is dominated by In-$s$ states with a little contribution from the anion $p$ states. The hybridization between the cation and anion orbitals in the occupied states indicating a strong covalent bond is present between them which is well known for the metal chalcogenides. Due this covalent bonding broad band features are found at band edges as shown in Fig.3.

In order to identify the suitability of InSe monolayer for efficient photocatalytic activity, it is necessary to align the band edge positions of VBM and CBM with respect to the redox potentials. Here, we adopt the commonly used value for oxidation potential and the reduction potential of water at 25°C are -5.67 eV and -4.44 eV, respectively, with respect to the absolute vacuum level. Fig. 4 shows the band edge alignment of InSe and Te doped InSe monolayers with the redox potentials. It is to be noted that the CBM of pristine InSe monolayer is located close to the hydrogen reduction potential ($H^+/H_2$) and the VBM is located more positively than the water oxidation potential ($O_2/H_2O$). The difference between the hydrogen reduction potential and the CBM edge is known as reducing power which is found to be 0.09 eV while the oxidizing power is the difference between the VBM and water oxidation potential is found to be 0.82 eV. This small reducing power to produce $H_2$ may diminish when strain is applied in any direction. Hence, we need to incorporate a co-catalyst to achieve the whole water splitting even under small strain.

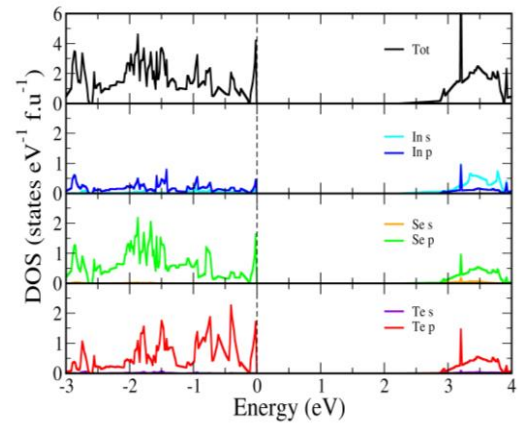

**FIGURE 3.** Calculated density of states for Te doped InSe by HSE06 functional. The Fermi level is set to zero.

The CBM of InSe monolayer needs to be shifted upwards while slightly stabilizing the VBM to improve its catalyzing efficiency. Hence, we have expected that atom with bigger size doping would be a promising strategy. Doping Te at Se site would give strain in the lattice giving elongation of 5.9 %. So we have substituted one Te for Se in InSe monolayer through a 2x2 supercell and this would correspond to a doping concentration of 6.25 wt. %. From the calculated band alignment one can observe that CBM of Te doped InSe shifted upwards a bit and its reducing power increased to 0.19 eV but the VBM position doesn't change. This shift in CBM indicate that a better photocatalyzing ability would be expected in Te doped InSe monolayers.

It may be noted that apart from appropriate bandgap and band alignment, there is another important criteria for a photocatalyst is to absorb the visible region of the solar light for the improved photocatalytic efficiency. As we mentioned earlier, the calculated bandgap values of both the pristine and Te doped InSe fall under the visible part of the solar spectra. Hence, these materials could absorb visible light which is about 42% in the solar spectra. In consistent with the above, J. Lauth *et. al* [7] reported experimentally that ultrathin InSe layers exhibits a broad absorption peak in the visible region at around 560 nm (~ 2.2 eV). Hence InSe and Te doped InSe monolayer could absorb a significant amount of solar light. Therefore, Te doped InSe monolayer would be a suitable candidate for the efficient photocatalyst for water splitting.

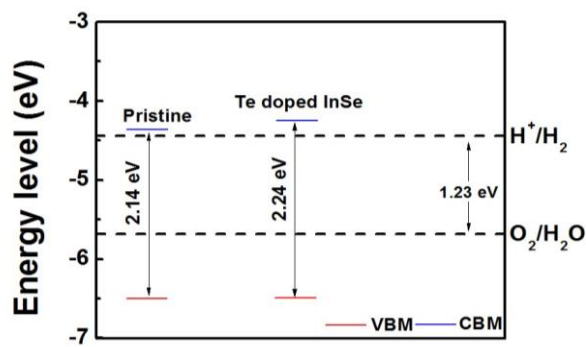

**FIGURE 4.** Band edge positions calculated by HSE06 functional. The dashed lines are standard water redox potentials. The reference potential is the vacuum level.

## CONCLUSION

In conclusion, we have performed density functional calculations to study the effect of Te on InSe monolayer for the photocatalytic water splitting. The calculated bandgaps are suitable for the photocatalytic water splitting and moreover it fall within the visible region of the solar spectrum indicating these materials could absorb a significant amount of solar light. Further, the density of states calculations illustrates the covalent bonding character between the atoms which is well known for the 2D layered system. The band alignment shows that Te doped InSe monolayer would straddle the redox potentials even under small strain for spontaneous water splitting under visible light. Our overall results indicate that Te substitution improves the photocatalytic water splitting ability in InSe monolayer and it could be a promising photocatalyst for water splitting.

## ACKNOWLEDGMENTS

The authors are grateful to the Research Council of Norway for computing time on the Norwegian supercomputer facilities. This research was supported by the Indo-Norwegian Cooperative Program (INCP) via Grant No. F.No. 58-12/2014(IC) and Department of Science and Technology, India via Grant No. SR/NM/NS-1123/2013.